\documentclass[a4paper,12pt]{article}
\usepackage{latexsym,amsmath,amsfonts,amssymb}
\usepackage[latin1]{inputenc}
\usepackage{moreverb,graphicx,color,caption,macros,multirow,footmisc}
\usepackage[numbers]{natbib}
\usepackage[pdftex]{hyperref}
\hypersetup{colorlinks,linkcolor=black,filecolor=black,urlcolor=black,citecolor=black,plainpages=false,
hypertexnames=false}

\begin{document}
\begin{flushright} HU-EP-09/59\\ SFB/CPP-09-114\\ DESY 09-203\end{flushright}
\begin{center}
  {\Large\bf  Symanzik improvement of lattice QCD with four f\/lavors of Wilson quarks}
\end{center}
\begin{figure}[!h]
$$\includegraphics[height=1cm]{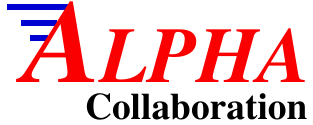}$$
\end{figure}
\begin{center}
Fatih Tekin\footnote{fatih@physik.hu-berlin.de\label{Fatih}}, Rainer Sommer\footnote{rainer.sommer@desy.de\label{Rainer}}, Ulli Wolf\/f\footnote{uwolff@physik.hu-berlin.de\label{Ulli}}
\vskip 0.5 cm
$^{\rm \ref{Fatih},\ref{Ulli}}$ 
Institut f\"ur Physik, Humboldt Universit\"at\\
Newtonstr. 15, 12489 Berlin,\\ 
Germany
\vskip 1.5ex
$^{\rm \ref{Rainer}}$
NIC, DESY\\
Platanenallee 6, 15738 Zeuthen,\\ 
Germany
\vskip 0.5 cm
{\bf Abstract}
\vskip 0.7ex
\end{center}
We have determined the non-perturbative $\mathcal{O}(a)$-improvement
coefficient $\csw$ for four f\/lavors of Wilson quarks with the plaquette gauge
action in a range of $\beta\geq 5.0$. 
The data are fitted with several Pad\'e approximation formulae to get an impression of the stability. A small extrapolation below $\beta=5.0$ seems acceptable.
\thispagestyle{empty}
\newpage
\section{Introduction}
The lattice regularization of QCD is a powerful tool to non-perturbatively study QCD in the low energy region. The numerical implementation involves however
a finite lattice spacing $a$ which has to be removed in the continuum limit. The rate of approaching the continuum limit will depend on the details of the lattice formulation. A systematic way to reduce the discretization effects order by order in $a$ is the Symanzik improvement programme \cite{Symanzik:1981hc,Symanzik:1983dc,Symanzik:1983gh} for on-shell quantities \cite{Luscher:1985zq,Luscher:1984xn}. In the case of Wilson fermions, Sheikoleslami and Wohlert \cite{Sheikholeslami:1985} have shown that for reducing the lattice artefacts from $\mathcal{O}(a)$ to $\mathcal{O}(a^2)$ only one additional dimension five operator in the Lagrangian is needed. To achieve this acceleration of the continuum limit non-perturbatively the coefficient $\csw$ of the corresponding operator has to be determined in numerical simulations. In the quenched case \cite{Luscher:1996ug}, the ALPHA collaboration has found \cite{Luscher:1992an,Sint:1993un,Sint:1995rb} that the non-perturbative result for $\csw$ deviates significantly from the one-loop perturbative value \cite{Sheikholeslami:1985,Wohlert:1987}. Furthermore, the effect of two species of dynamical fermions on $\csw$ was also studied by the ALPHA collaboration and the dif\/ference to the quenched case was clearly visible \cite{Jansen:1998mx}. The effect of a third flavor was studied by the CP-PACS and JLQCD collaborations \cite{Yamada:2004ja} with the result that $\csw (g_{0}^2)$ is not very much affected by it and their result is very close to the two flavor values of the ALPHA collaboration which can be seen in \Fig{cswall}. Our aim in this paper is to calculate $\csw$ non-perturbatively for four f\/lavors in the Schr\"odinger functional scheme. 

The paper is organized as follows. First we want to give a brief reminder of $\mathcal{O}(a)$ improved lattice QCD and briefly discuss the improvement condition for $\csw$. After discussing some features of the algorithmic implementation we will give a summary about the simulation parameters and the raw data. The procedure how to determine $\csw$ from the raw data will be described next. Finally we will give the conclusions.

\section{Improvement condition}
Our starting point for $\mathcal{O}(a)$ improved lattice QCD is the fermion action 
\begin{equation}
 S_{\text{f}}=a^4\sum_{x}\psibar(x)\left[D+m_{0}\right]\psi(x),
\end{equation}
where $a$ is the lattice spacing and $m_{0}$ is the bare quark mass. The matrix $D$ is the Wilson-Dirac operator \cite{Wilson:1974}
\begin{equation}
 D=\frac{1}{2}\left[\left(\nabstar\mu+\nab\mu\right)\dirac\mu-a\nabstar\mu\nab\mu\right]
\end{equation}
with the lattice covariant forward and backward derivatives $\nabla_{\mu}$ and $\nabla^{*}_{\mu}$. The leading order lattice artefacts in on-shell quantities which are calculated with this action are linear in $a$. However, the leading order discretization effects may be canceled by adding the so-called Sheikoleslami Wohlert term to the action \cite{Sheikholeslami:1985}
\begin{equation}\label{SWterm}
 D_{\rm improved}=D+\csw\frac{ia}{4}\sigma_{\mu\nu}\mathcal{F}_{\mu\nu}.
\end{equation}
The lattice field strength tensor $\mathcal{F}_{\mu\nu}$ is defined as in \cite{Luscher:1996sc} and $\sigma_{\mu\nu}=\frac{i}{2}[\dirac\mu,\dirac\nu]$. The coefficient $\csw$ in \eq{SWterm} is a function of the bare coupling $g_{0}$ and if it is chosen properly, $D_{\text{improved}}$ becomes the on-shell $\mathcal{O}(a)$ improved lattice Wilson-Dirac operator. 
For a complete cancellation of the $\mathcal{O}(a)$ effects\footnote{
This refers to the massless theory which we consider in connection with the \SF renormalization scheme.
}
in correlation functions, the local composite fields that enter also have to be improved \cite{Luscher:1996sc}. In our case, such composite fields are the isovector axial current $A_{\mu}^{a}(x)$ and the pseudo-scalar density $P^{a}(x)$. It turns out that only the isovector axial current needs an improvement because there is no dimension four operator with the same behavior as $P^{a}$ under the symmetries of the lattice theory. The $\mathcal{O}(a)$ improvement requires the combination
\begin{equation}
 (\aimpr)^{a}_{\mu}=A^{a}_{\mu}+a\cdot \ca\frac{1}{2}(\drvstar\mu+\drv\mu)P^{a}
\end{equation}
 where $A^{a}_{\mu}$ and $P^{a}$ are given by
\begin{eqnarray}
 A^{a}_{\mu}(x)&=&\psibar(x)\dirac\mu\dirac5\frac{\tau^{a}}{2}\psi(x)\\
 P^{a}(x)&=&\psibar(x)\dirac5\frac{\tau^{a}}{2}\psi(x),
\end{eqnarray}
$\drv\mu$, $\drvstar\mu$ are the forward and backward dif\/ference operators and $\tau^{a}$ are Pauli matrices acting on one pair among the four degenerate f\/lavors. The improvement coefficient $\ca$ is known in perturbation theory \cite{Luscher:1996vw} and from non-perturbative determinations for $\Nf=0$ \cite{Luscher:1996ug} and $\Nf=2$ \cite{DellaMorte:2005se,DellaMorte:2008xb}. 
As explained in \cite{Luscher:1996ug,Jansen:1998mx}, we introduce the unrenormalized PCAC quark mass
\begin{equation}\label{mpcac}
 m(x_{0})=\frac{\frac{1}{2}(\drv0+\drv0)\fa(x_{0})+\ca a\drvstar0\drv0\fp(x_{0})}{2\fp(x_{0})}
\end{equation}
where the correlation functions $\fa$ and $\fp$ contain $A^{a}_{\mu}$ and $P^{a}$ and are given by (2.1) and (2.2) of \cite{Luscher:1996ug}. A second mass $m'$ can be defined in the same way as \eq{mpcac} but with the primed correlation functions $\faprime$ and $\fpprime$ ((2.5) and (2.6) of \cite{Luscher:1996ug}). The unprimed and primed correlation functions are related to each other by a time reflection in the Schr\"odinger functional. Since the boundary conditions
\begin{eqnarray}
 U(x,k)|_{x_{0}=0}&=&\exp\{aC_{k}\};\quad C_{k}=\frac{i}{6L}\text{diag}(-\pi,0,\pi)\\
 U(x,k)|_{x_{0}=T}&=&\exp\{aC_{k}'\};\quad C_{k}'=\frac{i}{6L}\text{diag}(-5\pi,2\pi,3\pi)
\end{eqnarray}
are such that $C_{k}$ and $C_{k}'$ are not the same, $\fx$ and $\fxprime$ also dif\/fer. 
Since with PCAC we have, however, inserted an `operator identity', all $m(x_0), m'(y_0)$ dif\/fer only at the level of
$\mathcal{O}(a^2)$ effects in the improved theory. We could hence
for some choice, such as $x_{0}=y_0=\frac{T}{2}$, impose $m-m'=0$ as one condition for the proper choice of $\csw$ and $\ca$. Because the coefficient $\ca$ is a priori not known it is advantageous however to first eliminate this parameter and define a quark mass $M$ independent of $\ca$ which agrees with the quark mass $m$ up to $\mathcal{O}(a^2)$ effects. For this purpose we name
the partial contributions
\begin{eqnarray}\label{r(x)}
r(x_{0})&=&\frac{(\drvstar0+\drv0)\fa(x_{0})}{4\fp(x_{0})}\\\label{s(x)}
s(x_{0})&=&\frac{a\drvstar0\drv0\fp(x_{0})}{2\fp(x_{0})}
\end{eqnarray}
and rewrite the mass $m$ as
\begin{equation}
m(x_{0})=r(x_{0})+\ca s(x_{0}).
\end{equation}
With an analogous definition for $m'$, a quark mass $M$ can than be written in the following way 
\begin{equation}
M(x_{0},y_{0})=m(x_{0})-s(x_{0})\frac{m(y_{0})-m^{\prime}(y_{0})}{s(y_{0})-s^{\prime}(y_{0})}=r(x_{0})-s(x_{0})\frac{r(y_{0})-r^{\prime}(y_{0})}{s(y_{0})-s^{\prime}(y_{0})}.
\end{equation}
In this combination, which in the improved theory dif\/fers from $m$ by $\mathcal{O}(a^2)$ only, $\ca$ drops out.
Now, we define $M^{\prime}(x_{0},y_{0})$ analogously and could require that the dif\/ference
\begin{equation}
\Delta M\left(\frac{3}{4}T,\frac{1}{4}T\right)=M\left(\frac{3}{4}T,\frac{1}{4}T\right)-M^{\prime}\left(\frac{3}{4}T,\frac{1}{4}T\right)
\end{equation}
has to vanishes for our value of $\csw$. The choice $(x_{0},y_{0})=(\frac{3}{4}T,\frac{1}{4}T)$ is one possible choice \cite{Luscher:1996ug} for the argument of $\Delta M$. For the quark mass $M$ itself we choose $(x_{0},y_{0})=(\frac{1}{2}T,\frac{1}{4}T)$ \cite{Luscher:1996ug}. In order to reproduce the tree level value of $\csw$ exactly
for finite $a$, 
we finally impose the improvement condition
\begin{equation}\label{treelevel}
 \Delta M=\Delta M^{(0)}
\end{equation}
where $\Delta M^{(0)}$ is the tree level value of perturbation theory in the $\mathcal{O}(a)$ improved theory. For $L/a=8$ one finds for example \cite{Luscher:1996ug}
 \begin{equation}\label{Mtreelevel}
 a\Delta M^{(0)}|_{M=0,\csw=1}=0.000277.
\end{equation}
As discussed in \cite{Sommer:2006sj}, improvement coefficients possess a unique perturbative expansion
but non-perturbatively they are themselves ambiguous by cutoff terms and thus depend on the choice of the improvement conditions.
Of course, for QCD these uncertainties amount to cutoff effects beyond the order that is improved, $\mathcal{O}(a^2)$ in the case at hand.
In principle one then has to determine improvement coefficients for one fixed set of conditions
as functions of $g_{0}$ at constant physics, i.~e. fixing all scale ratios except $a$ that shrinks with $g_{0}$ .
For the Schr\"odinger functional
this would in particular require constant $L/r_0$ as $g_0$ is lowered.
For practical reasons we fix however $a/L$ instead and refer the reader to sect. (I.2.4.1) of \cite{Sommer:2006sj}
for a detailed discussion. The replacement of zero by the small tree-level value on the right hand side of
\eq{Mtreelevel} guarantees that our definition has the correct limit for $g_0 \to 0$.
\section{Simulations}
Our simulations for $\Nf=4$ are based on an adaptation of TAO codes ---
suitable for APE computers \cite{Alfieri:2001xm} --- used earlier by the ALPHA collaboration
for $\Nf=2$ studies. An ordinary HMC algorithm \cite{Duane1987216} has been implemented with symmetric even-odd preconditioning \cite{Gupta:1989, Degrand:1990, Witzel:2008}
and the Sexton-Weingarten integration scheme \cite{Sexton:1992}. Mass preconditioning \cite{Hasenbusch:2001ne,Hasenbusch:2002ai} was not enforced
as we expect that the gain for the Schr\"odinger functional with the parameters envisaged here 
would  not be so significant \cite{DellaMorte:2003jj}. For the sake of convenience, from here on, we will set the lattice spacing $a$ to one.

\subsection{Algorithm}
The fermionic determinant for an \textit{even} number of f\/lavors can be represented by pseudo-fermion fields $\phi_i$, $\phi_i^{\dagger}, i=1,\ldots,\Nf/2$.
The partition function then reads for $\Nf=4$
\begin{equation}
Z=\int DU\,D\phi_1^{\dagger}\,D\phi_1\,D\phi_2^{\dagger}\,D\phi_2\,\, {\rm e}^{-\Sg-\sum\nolimits_{i=1,2}\phi_i^{\dagger}(QQ^{\dagger})^{-1}\phi_i}
\end{equation}
where $\Sg$ is Wilson's plaquette gauge action \cite{Wilson:1974}
and $Q$ is related to the Dirac matrix $M$ by
\begin{equation}\label{evenoddM}
 Q=\dirac5 M =\dirac5 
 \left(
    \begin{array}{cc}
      M_{\rm ee}&M_{\rm eo}\\M_{\rm oe}&M_{\rm oo}
    \end{array}
 \right).
\end{equation}
The key idea of the even-odd preconditioning \cite{Gupta:1989, Degrand:1990} is to divide the lattice sites into even and odd sites according to the sum over the coordinates $\sum_{\mu=0}^3 x_{\mu}$. If it is \textit{even} for the lattice site $x$, the site is called \textit{even} otherwise it is called \textit{odd}. Following this strategy, the Dirac matrix $M$ decomposes into the block form in \eq{evenoddM}. The components are given as follows \cite{Jansen:1996yt}
\begin{multline}
M_{x,x'}=(1+T_{x,x})\delta_{x,x'}\\
-\kappa\sum_{\mu}\left[(1-\dirac\mu)U_{\mu}(x)\delta_{x+\hat\mu,x'}+(1+\dirac\mu)U_{\mu}^{\dagger}(x-\hat\mu)\delta_{x-\hat\mu,x'}\right].
\end{multline}
The matrices $T_{x,x}$ 
\begin{equation}
T_{x,x}=\frac{i}{2}\csw\kappa\sigma_{\mu\nu}\mathcal{F}_{\mu\nu}(x)
\end{equation}
vanish if the improvement coefficient $\csw$ is set to zero and the submatrices $M_{\rm ee}$ and $M_{\rm oo}$ then become equal to the unit matrix. In our case however, we consider the $\mathcal{O}(a)$ improved Sheikoleslami-Wohlert action where the coefficient $\csw$ is determined non-perturbatively for $\Nf=4$. Furthermore, the submatrices of $M$ possess the following properties
\begin{equation}
\begin{array}{rclp{2cm}rcl}
M_{\rm ee}^{\dagger} & = & M_{\rm ee} & & M_{\rm oo}^{\dagger} & = & M_{\rm oo} \\
M_{\rm eo}^{\dagger} & = & \dirac5 M_{\rm oe}\dirac5 & & M_{\rm oe}^{\dagger} & = & \dirac5 M_{\rm eo}\dirac5.
\end{array}
\end{equation}
The origin of the algorithmic acceleration by even-odd preconditioning lies in
the factorization of the determinant. In this context, two possibilities
appear. We may either factorize out only $M_{\rm ee}$ or $M_{\rm oo}$
(\textit{asymmetric}), or second, we extract both factors
(\textit{symmetric}). The derivation of both versions is very similar. The key
point is the calculation of a determinant of matrices like \eq{evenoddM} using
its \textit{Schur complement}. For a general matrix which consist of
submatrices, the determinant can be calculated in the following way 
\begin{equation}\label{Schur}
\det\left(\begin{array}{cc}A&B\\C&D\end{array}\right)=\det\{AD-ACA^{-1}B\}.
\end{equation}
In our case, the determinant of $Q=\dirac5M$ leads to
\begin{equation}
\det Q=\det\{M_{\rm ee}\}\det\{M_{\rm oo}\}\det\{\hat{Q}\}
\end{equation}
for the symmetric even-odd preconditioning, where
\begin{equation}\label{Qsymm}
\hat{Q}=\dirac5(1-M_{\rm oo}^{-1}M_{\rm oe}M_{\rm ee}^{-1}M_{\rm eo})\,.
\end{equation}
In our numerical implementation, we
only use the above described symmetric even-odd preconditioning but there is
no fundamental problem to implement the asymmetric even-odd preconditioned
version. However, it should be kept in mind that the authors of
\cite{Aoki:2002} found that the performance of HMC with symmetric even-odd
preconditioning is roughly 30\% better
than the HMC algorithm with asymmetric even-odd preconditioning. Due to the decomposition in \eq{Qsymm}, only the odd components $\phi_{o}$ of the fields $\phi$ appear.
With symmetric preconditioning the partition function for four f\/lavors now reads
\begin{equation}
Z=\int DU\,D\phi_{{\rm o}1}^{\dagger}\,D\phi_{{\rm o}1}\,D\phi_{{\rm o}2}^{\dagger}\,D\phi_{{\rm o}2}\,\,{\rm e}^{-\Sg[U]-S_{\rm det}[U]-S_{\rm pf}[U,\phi^{\dagger}_{{\rm o}i},\phi_{{\rm o}i}]}
\end{equation}
with the gauge part $\Sg[U]$ 
and
\begin{eqnarray}\label{detpart}
S_{\rm det}&=&\Nf\left[\ln\det\{M_{\rm ee}\}+\ln\det\{M_{\rm oo}\}\right]\\\label{pfpart}
S_{\rm pf}&=&\sum\nolimits_{i=1,2}\phi_{{\rm o}i}^{\dagger}(\hat{Q}\hat{Q}^{\dagger})^{-1}\phi_{{\rm o}i}.
\end{eqnarray}

\subsection{Simulation parameters and raw results}
The simulations were performed in the Schr\"odinger functional scheme
\cite{Luscher:1992an,Sint:1993un,Sint:1995rb} with periodic boundary
conditions for the spatial extension and Dirichlet boundary conditions in the
temporal direction. This means the phase $\theta$ was here set to zero in all
runs. The data were obtained on hypercubic Euclidean $16\times 8^3$
lattices. The $\mathcal{O}(a)$ improvement of the \SF requires additional 
improvement terms at the boundaries. However, since PCAC is an
operator relation, these terms are irrelevant
for a correct determination of $\csw$. Nevertheless, we have chosen them 
as follows. The pure gauge part of the action acquires a weight 
$w(p)=\ct(g_{0})$ of time-like plaquettes $p$ attached to the boundary planes
for which we have inserted the 2-loop value \cite{Bode:1999sm}
\begin{multline}
 \ct(g_{0})=1+[-0.08900+0.0191410\Nf]g_{0}^{2}\\+[-0.0294+0.002\Nf+0\Nf^2]g_{0}^{4}\,,
\end{multline}
and for the fermionic improvement coefficient 
$\ctildet(g_{0})$ \cite{Luscher:1996vw} 
we take the 1-loop perturbative value \cite{Sint:1997jx}
\begin{equation}
 \ctildet(g_{0})=1-0.01795g_{0}^{2}\,.
\end{equation}
In all our simulations the trajectory length was kept fixed to one. We chose the intervals in $\beta$ similar to \cite{Jansen:1998mx}. The mean acceptance rate was around $90\%$ and we performed 4400 trajectories per value of $\beta$ and $\csw$ on average. A large part of our computations ran on APEmille machines with 128 processors each. For some values of $\beta$ and $\csw$ we also used apeNEXT crates with 256 processors. The summary table of the measurements can be found in the appendix.

\section{Determination of \texorpdfstring{$\csw$}{}}
\begin{figure}[htb]
\begin{center}
$$\includegraphics[width=10cm,height=4cm]{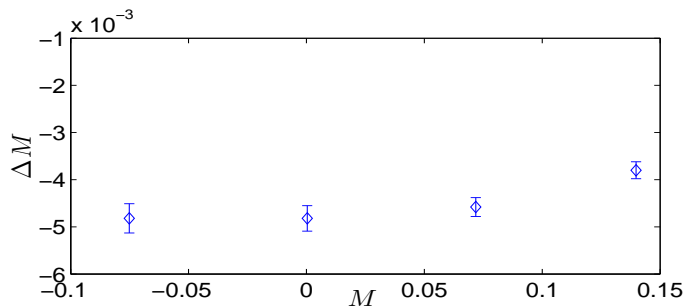}$$
\caption{Mass dependence of $\Delta M$ at $\beta=5.0$ and $\csw=2.4$. The errors of $M$ are smaller than the symbol sizes.}\label{dependence}
\end{center}
\end{figure}

\begin{figure}[htb]
$$\includegraphics[width=10cm]{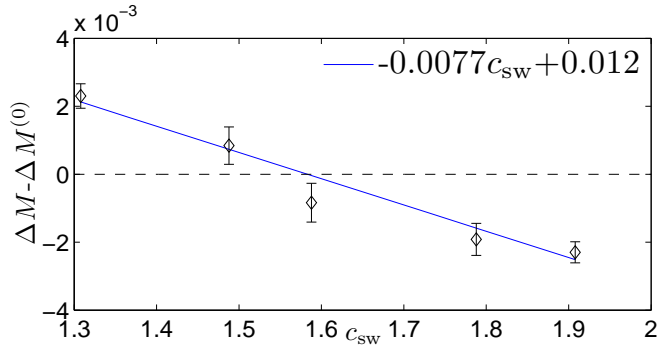}$$
\caption{Determination of $\csw$ at $\beta=5.4$. The desired value $\csw^{*}$ is located at the point where the fit curve (solid) passes through zero (dashed).}\label{cswstar}
\end{figure}

The numerical procedure for the determination of $\csw$ involves the following  main steps
\begin{enumerate}
 \item\label{first} Compute $\Delta M$ and $M$ for several $\kappa$ at fixed $g_{0}^2$ (or $\beta$) and $\csw$. Then interpolate linearly in $M$ to find $\Delta M$ at vanishing quark mass $M=0$.
 \item For fixed $g_{0}^{2}$, repeat step \ref{first} for several values of $\csw$ and find $\csw^{*}$ which solves \eq{treelevel} by a linear fit in $\csw$.
 \item Repeat the preceding steps for a suf\/ficient range of $g_{0}^2$ and fit these data with an appropriate function to represent the smooth functional dependence of $\csw$ on $g_{0}^2$.
\end{enumerate}
This procedure is computer time demanding because for each value of $\beta$ and for each value of $\csw$ of \Tab{table}, we would need several runs, at least three, for interpolating $\Delta M$ in $M$ to $M=0$. To save computer time, we modified this method slightly. As discussed in the determination of $\csw$ for $\Nf=0,2$ \cite{Luscher:1996ug,Jansen:1998mx}, the weak dependence of $\Delta M$ on $M$ holds also at $\Nf=4$. For one set of parameters $\beta$ and $\csw$, we checked the dependence explicitly (\fig{dependence}).
Since $\Delta M$ depends weakly on $M$, we contented ourselves with determining $\Delta M$ for some  $|M|< 0.03$ \cite{Jansen:1998mx} and used these values of $\Delta M$ as an approximation for $\Delta M$ at $M=0$. A typical result is shown in \Fig{cswstar}.
We performed simulations from $\beta=12$ to $\beta=5.0$. For each $\beta$, we
calculated the observables $\Delta M$ and $M$ at least for three dif\/ferent
values of $\csw$ in such a way that the condition $|M|<0.03$ held and that
$\Delta M$ had a change of sign. The linear interpolation 
$\Delta M = s\;(\csw-\csw^*) + \Delta M^{(0)}$ yields the
desired values $\csw^*$ shown in \Tab{interpolation}.
\begin{table}[htb]
$$
\begin{tabular}{|l|l||l|l|}\hline
$\beta$ & $\csw^{*}$ & $\beta$ & $\csw^{*}$\\\hline\hline
12    & 1.1429(39) & 6.0   & 1.463(19) \\\hline
9.6   & 1.1895(62) & 5.7   & 1.554(17) \\\hline
7.4   & 1.2955(76) & 5.4   & 1.583(25) \\\hline
6.8   & 1.3375(94) & 5.2   & 1.614(28) \\\hline
6.3   & 1.389(12)  & 5.0   & 1.717(31) \\\hline
\end{tabular}
$$
\caption{Results of the linear interpolation}\label{interpolation}
\end{table}
We also tried to go below $\beta=5.0$ but at $\beta=4.8$, we were not able to
locate a significant sign change of $\Delta M$ in our data and therefore after
some attempts, we decided to stop searching. 

The CP-PACS and JLQCD collaborations computed $\csw$ for $\Nf=3$ in the
Schr\"odinger functional setup of lattice QCD with the plaquette gauge 
action \cite{Yamada:2004ja}. They found that the result for three f\/lavors is 
very close to the two flavor result \cite{Jansen:1998mx}. 
In addition they calculated $\csw$ with four f\/lavors for $\beta=9.6$ 
and found $\csw^* = 1.1954(48)$  in good agreement with our 
$\csw^* =  1.1895(62)$ .

\begin{figure}[htb]
$$\includegraphics[width=10cm,height=6.5cm]{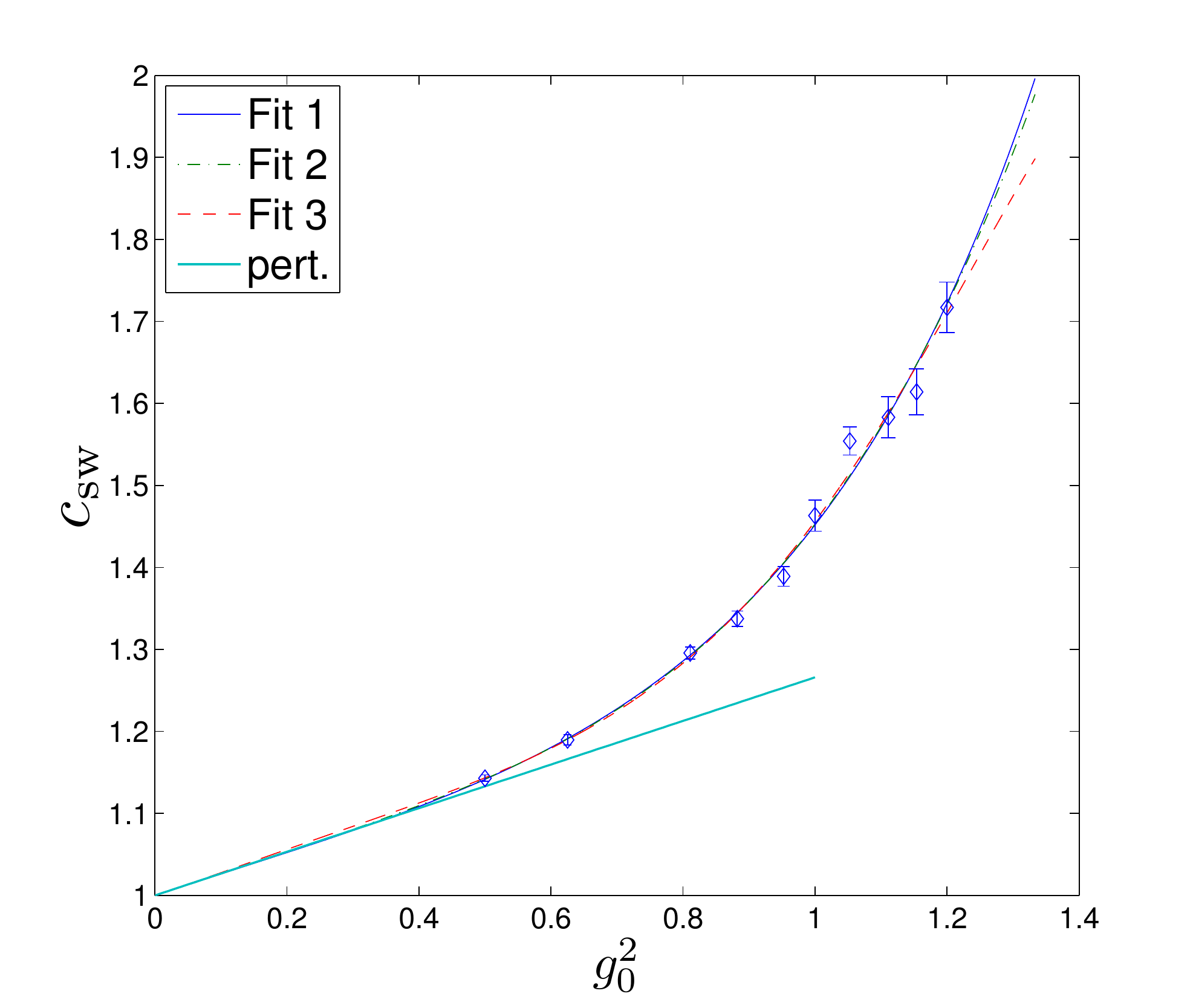}$$
\caption{Comparison of dif\/ferent Pad\'e-approximation formula for our data.}\label{fitall}
\end{figure}

After obtaining the proper values $\csw^{*}$ which satisfy the improvement condition \eq{treelevel},
we want to represent and interpolate our data by a simple Pad\'e formula, appropriate for the achieved precision,
which also incorporates the known 1-loop perturbative result.
The solution that we want to advocate here for $\Nf=4$ is
\begin{equation}\label{mainresult}
\csw(g_{0}^{2})=\frac{1-0.1372g_{0}^2-0.1641g_{0}^4+0.1679g_{0}^6}{1-0.4031g_{0}^2}\quad0\le g_{0}^{2}\le 1.2.
\end{equation}
This curve appears as {\tt Fit1} in \Fig{fitall}.
The two other lines {\tt Fit2} and {\tt Fit3} include one and two more powers in the numerator.
The deviation of the dif\/ferent fit formulae in the range $g_{0}^{2}\in[0,1.2]$ ($\beta=[12.0,5.0]$) is negligible and beyond $g_{0}^{2}=1.2$, {\tt Fit1} and {\tt Fit2} are almost the same down to $\beta=4.5$ but {\tt Fit3} deviates slightly. Our non-perturbatively determined formula \eq{mainresult} for $\csw$ with four f\/lavors is valid down to $\beta=5.0$ ($g_{0}^{2}=1.2$) but may perhaps be used to $\beta=4.5$ within a small uncertainty.
 $$\includegraphics[width=10cm,height=6.5cm]{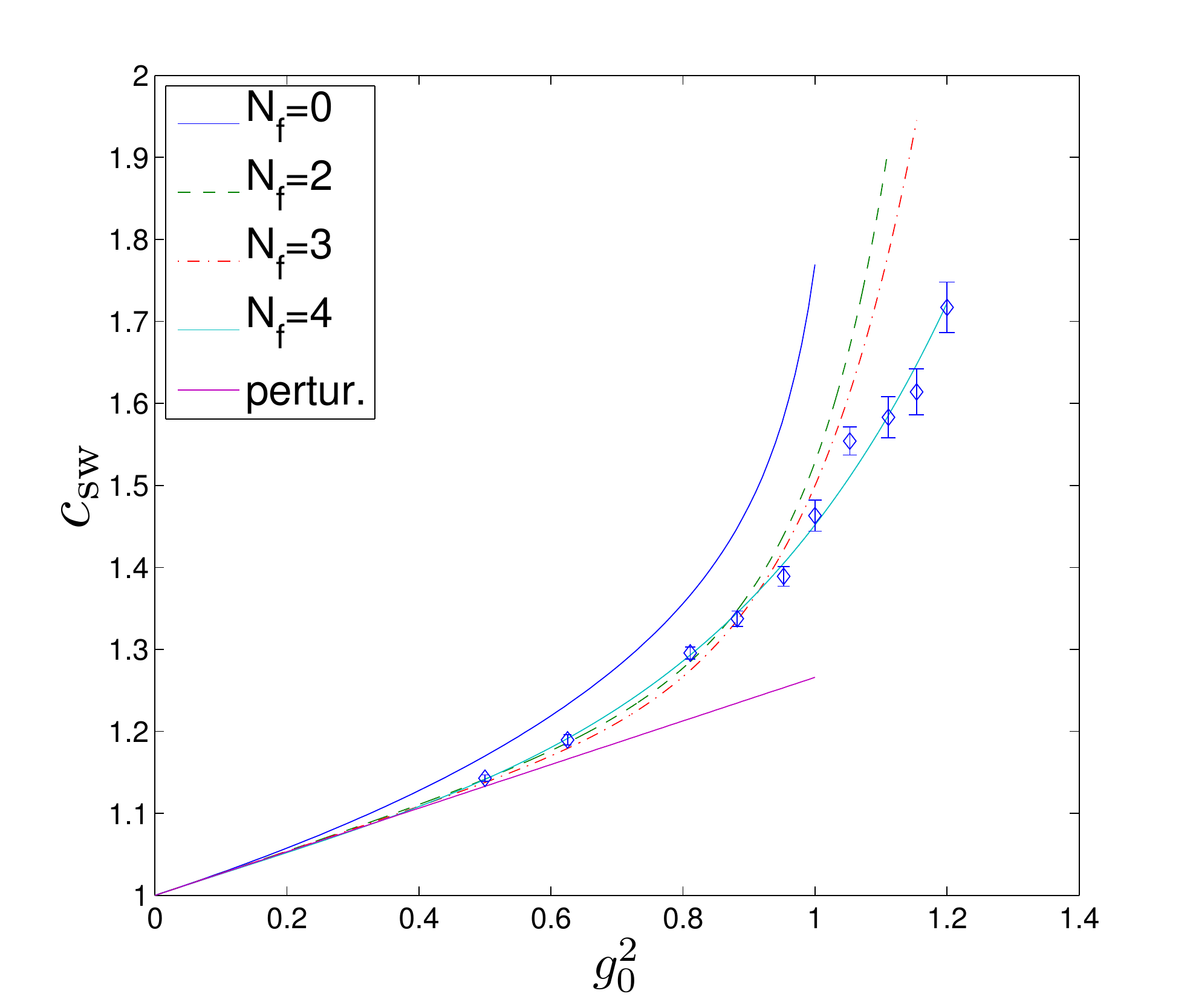}$$
 \captionof{figure}{Summary plot of all known $\csw(g_{0}^{2},\Nf)$ for the plaquette gauge action.}\label{cswall}
\vspace{0.75cm}
To conclude in \Fig{cswall}, we juxtapose our new data and fit formula at $\Nf=4$ to those known for $\Nf=0,2,3$.

\section{Conclusions}
In the present paper, we have performed simulations for the calculation of the
improvement coefficient $\csw$ for four f\/lavors of Wilson fermions in a range
of $\beta\geq5.0$ and give \eq{mainresult} as a suitable parameterization of
our data. We compared three simple-minded dif\/ferent Pad\'e-approximation
formulae for our data and could show that a small extrapolation of $\csw$
beyond $\beta=5.0$ to $\beta=4.5$ with the main formula \eq{mainresult} may
still be acceptable.
An immediate use of $\csw$ will be the determination of the 
non-perturbative running 
of the \SF coupling for $\nf=4$ massless f\/lavors, extending the present
knowledge which comprises $\nf=0$ \cite{Luscher:1996ug}, $\nf=2$ \cite{Jansen:1998mx} 
and $\nf=3$ \cite{Yamada:2004ja}.

\section*{Acknowledgements}
This work is part of the ALPHA-collaboration research program. We thank NIC
for allocating computer time on the APE computers at DESY, Zeuthen and the
staf\/f of the computer center at Zeuthen for their support. Fatih Tekin thanks Oliver Witzel for numerous fruitful discussions about the details of the ALPHA code and for useful suggestions. This work is
supported by the Deutsche Forschungsgemeinschaft (DFG) in the framework of
SFB/TR 09 and by the European Community
through EU Contract No.~MRTN-CT-2006-035482, ``FLAVIAnet''.
\appendix
\section*{Appendix}
Table of data:
{\small
$$
\begin{array}{|l|l|l|r|r|}\hline
\beta                 & \kappa   & c_{sw}   &     M         & \Delta M    \\\hline\hline
 \multirow{3}{*}{12}  & 0.130280 & 1.028654 &  0.000224(75) &  0.001735(89)\\
                      & 0.129897 & 1.128654 & -0.001510(74) &  0.000468(84)\\
                      & 0.129449 & 1.228654 & -0.001371(82) & -0.000827(86)\\\hline
 \multirow{3}{*}{9.6} & 0.131516 & 1.140488 & -0.00004(11)  &  0.00077(12) \\
                      & 0.131164 & 1.170488 &  0.00595(11)  &  0.00053(11) \\
                      & 0.131164 & 1.250488 & -0.00853(11)  & -0.00040(12) \\\hline
 \multirow{4}{*}{7.4} & 0.134626 & 1.163222 & -0.00196(20)  &  0.00169(19) \\
                      & 0.133753 & 1.263222 &  0.00072(20)  &  0.00069(15) \\
                      & 0.132989 & 1.363222 &  0.00018(20)  & -0.00036(16) \\
                      & 0.132349 & 1.443222 &  0.00027(20)  & -0.00162(17) \\\hline
 \multirow{5}{*}{6.8} & 0.135638 & 1.209613 &  0.00128(30)  &  0.00169(26) \\
                      & 0.135082 & 1.299613 & -0.00642(32)  &  0.00033(26) \\
                      & 0.134896 & 1.309613 & -0.00333(29)  &  0.00079(24) \\
                      & 0.133813 & 1.409613 &  0.00320(29)  & -0.00000(24) \\
                      & 0.133056 & 1.509613 & -0.00141(30)  & -0.00204(22) \\\hline
 \multirow{4}{*}{6.3} & 0.137098 & 1.239058 & -0.00129(38)  &  0.00199(24) \\
                      & 0.136018 & 1.339058 &  0.00142(36)  &  0.00077(27) \\
                      & 0.135028 & 1.439058 &  0.00140(36)  & -0.00075(27) \\
                      & 0.134028 & 1.550580 & -0.00251(31)  & -0.00115(24) \\\hline
 \multirow{3}{*}{6.0} & 0.138358 & 1.250000 & -0.00090(85)  &  0.00270(30) \\
                      & 0.136669 & 1.387912 &  0.00214(47)  &  0.00022(48) \\
                      & 0.134375 & 1.587912 &  0.00677(43)  & -0.00123(34) \\\hline
 \multirow{5}{*}{5.7} & 0.140327 & 1.250000 & -0.0082(16)   &  0.00369(38) \\
                      & 0.138229 & 1.387912 &  0.00034(63)  &  0.00107(43) \\
                      & 0.137008 & 1.487912 &  0.00081(55)  &  0.00076(48) \\
                      & 0.135685 & 1.587912 &  0.00564(63)  & -0.00032(37) \\
                      & 0.133940 & 1.754350 & -0.00004(38)  & -0.00156(28) \\\hline
 \multirow{5}{*}{5.4} & 0.141417 & 1.307912 & -0.00032(77)  &  0.00258(36) \\
                      & 0.139111 & 1.487912 & -0.00712(89)  &  0.00112(55) \\
                      & 0.137815 & 1.587912 & -0.00762(70)  & -0.00056(57) \\
                      & 0.135028 & 1.787912 & -0.00117(69)  & -0.00164(47) \\
                      & 0.133775 & 1.907912 & -0.00854(39)  & -0.00202(31) \\\hline
 \multirow{6}{*}{5.2} & 0.143363 & 1.307912 & -0.0004(14)   &  0.00196(48) \\
                      & 0.140628 & 1.487912 & -0.00032(87)  &  0.00076(42) \\
                      & 0.139206 & 1.587912 & -0.00326(69)  &  0.00102(39) \\
                      & 0.138147 & 1.655891 &  0.00162(94)  &  0.00028(47) \\
                      & 0.136248 & 1.787912 &  0.00030(91)  & -0.00084(65) \\
                      & 0.134556 & 1.907912 &  0.00372(62)  & -0.00189(38) \\\hline
\end{array}
$$
$$
\begin{array}{|l|l|l|r|r|}\hline
 \multirow{6}{*}{5.0} & 0.146056 & 1.307912 &  0.0051(25)   &  0.00252(62) \\
                      & 0.142554 & 1.507912 &  0.0021(13)   &  0.00125(41) \\
                      & 0.138141 & 1.787912 & -0.0053(11)   &  0.00102(68) \\
                      & 0.136527 & 1.885463 &  0.0009(11)   &  0.00004(55) \\
                      & 0.135039 & 2.000000 & -0.00826(90)  & -0.00070(59) \\
                      & 0.129603 & 2.400000 &  0.00033(42)  & -0.00482(27) \\\hline
 \multirow{7}{*}{4.8} & 0.145928 & 1.500000 &  0.0083(54)   & -0.0002(12)  \\
                      & 0.142201 & 1.700000 & -0.0022(31)   & -0.0000(11)  \\
                      & 0.138295 & 1.910000 &  0.0017(13)   &  0.00060(62) \\
                      & 0.137971 & 1.930000 &  0.0031(12)   & -0.00054(48) \\
                      & 0.136844 & 2.000000 &  0.0008(11)   & -0.00044(50) \\
                      & 0.135327 & 2.100000 & -0.00492(87)  & -0.00116(44) \\
                      & 0.132358 & 2.300000 & -0.00550(72)  & -0.00244(41) \\\hline
\end{array}
$$
}
\captionof{table}{Summary table of the measurements}\label{table}
\bibliographystyle{unsrtnat}
\bibliography{bibliography}

\end{document}